# Multiple Power Quality Event Detection and Classification using Wavelet Transform and Random Forest Classifier

Sambit Dash
Department of Electrical Engineering
IIIT,Bhubaneswar
sambitdash.2011@gmail.com

Umamani Subudhi
Department of Electrical Engineering
IIIT,Bhubaneswar
umamani@iiit-bh.ac.in

***Abstract*—In this paper a technique for detection of multiple power quality (PQ) events is illustrated. An algorithm based on wavelet transform and Random Forest based classifier is proposed in this paper. The developed technique is implemented on 11 different power quality events consisting of single stage power quality events such as sag, swell, flicker, interruption and multi stage power quality events such as harmonics combined with sag, swell, flicker, interruption. PQ events are simulated in MATLAB using standard IEEE-1159 standard. Significant features of PQ events are extracted using wavelet transform and used to train random forest based classifier. The efficiency of Random Forest Based classifier is compared with other widely used machine learning algorithms such as K-Nearest Neighbour (KNN) and Support Vector Machine (SVM). From confusion matrix of different algorithms it is concluded that Random Forest shows superior classification accuracy as compared to SVM and KNN.**

***Keywords—Power Quality(PQ), wavelet transform, Random Forest, Classification***

## I. INTRODUCTION

The variations in voltage, current and frequency which affect electrical equipments constitute power quality phenomena. Depletion of natural resources such as coal and gas led to the advent of renewable energy producing sources such as wind and solar energy. But the renewable sources suffer from intermittency which causes disturbance on the electrical grid when such sources are integrated. Wind and solar energy systems use various types of converters to overcome their varying characteristics and improve efficiency but such converters are non linear in nature. Multiple household equipments are dc in nature due to which non linear current is drawn at multiple distribution sites which distort the voltage waveform at the point of common coupling. Poor power quality is particularly disastrous to industrial systems such as wood processing and semi-conductor based chip production industry[1]. PQ disturbances are defined as sudden changes in the shape and size of voltage waveform which have a beginning and end. If more than one event occur simultaneously then the events are called as multistage events. Such disturbances are caused due to non linear nature of load, ground faults, transient spike while switching heavy loads, sudden surge currents due to lightning strike.. Much of power quality disturbances are generated at the consumer end due to the usage of non linear loads and improper usage of electrical devices[2]. Such disturbances propagate to the transmission and distribution section of electrical grid causing dysfunction of electrical equipments along with extensive financial and economic losses. The ranges in PQ events which may damage or affect electrical equipments and change power quality are set by international standards such as IEEE-1159, IEC 61000, and EN 50160[3]. Such standards are crucial while designing PQ detection and mitigation devices. In the face of such problems it is quintessential to develop automatic power quality event detection and classification equipment which capture the faults at the earliest and alert the users of its occurance[4]. Extensive research into the area of power monitoring and power quality control along with mitigation of such issues using appropriate devices are the need of the hour. Multiple digital signal processing techniques have been used in the detection of power quality events chiefly among them are Fourier Transform(FT), Short Time Fourier Transform (STFT), wavelet transform. Usually power quality events are non stationary in nature thus Fourier Transform doesnot perform well for such utility. The disadvantage of fourier transform is its inability to provide any time domain localization of the events. Although the frequency components of the signal is detected but no information on the time at which the particular frequency component is present. To overcome the disadvantages of FT[5] Short Time fourier transform (STFT)[6] was introduced. STFT decomposes the non stationary signal into a series of stationary signals. Basically an appropriate window is chosen which divides the signal into multiple equal components such that in each window the frequency levels present are detected. But the efficiency of STFT depends heavily on the choice of window size. Improper selection of window size causes improper frequency locatization with respect to the given duration of time. To overcome the disadvantages of STFT, wavelet transform is introduced. Wavelet take care of the selection of window size and it gives proper frequency locatization with respect to the given time interval. After passing the signal through wavelet transform[7] multiple features from the wavelet transform are fed to a classifier then trained and tested on new signals. Due to Heisenberg's uncertainty principle it is not possible to detect frequency at each interval of time rather the frequency component within a specific interval of time can only be detected. Thus in this respect wavelet transform helps to localize the frequency components by diving the signal into multiple resolution. Practical signal generally made up of low frequency component of large duration and high frequency components of short duration. Utilizing this feature, wavelet transform decomposes the signal into approximate and detail coefficient .The approximate coefficient is the low frequency component and detail coefficient is high frequency component of the signal.

## II. Signal Processing Method

### A. *Discrete Wavelet Transform (DWT)*

In this paper the synthetic signal generated through MATLAB is fed to the wavelet transform and for mother wavelet, Daubechies4 (dB4) is selected as its performance is superior in detecting power quality disturbance. When a signal is fed to wavelet transform it is actually passed through multiple high pass and low pass filters. The filters decompose the voltage signal to approximate and detail coefficient. Approximate is high scale which gives the low frequency component present in the signal and detail is the low scale which gives the high frequency component of the signal. Multiple iterations of detail coefficient removes the low frequency component successively providing multiple lower resolution component of the same signal.

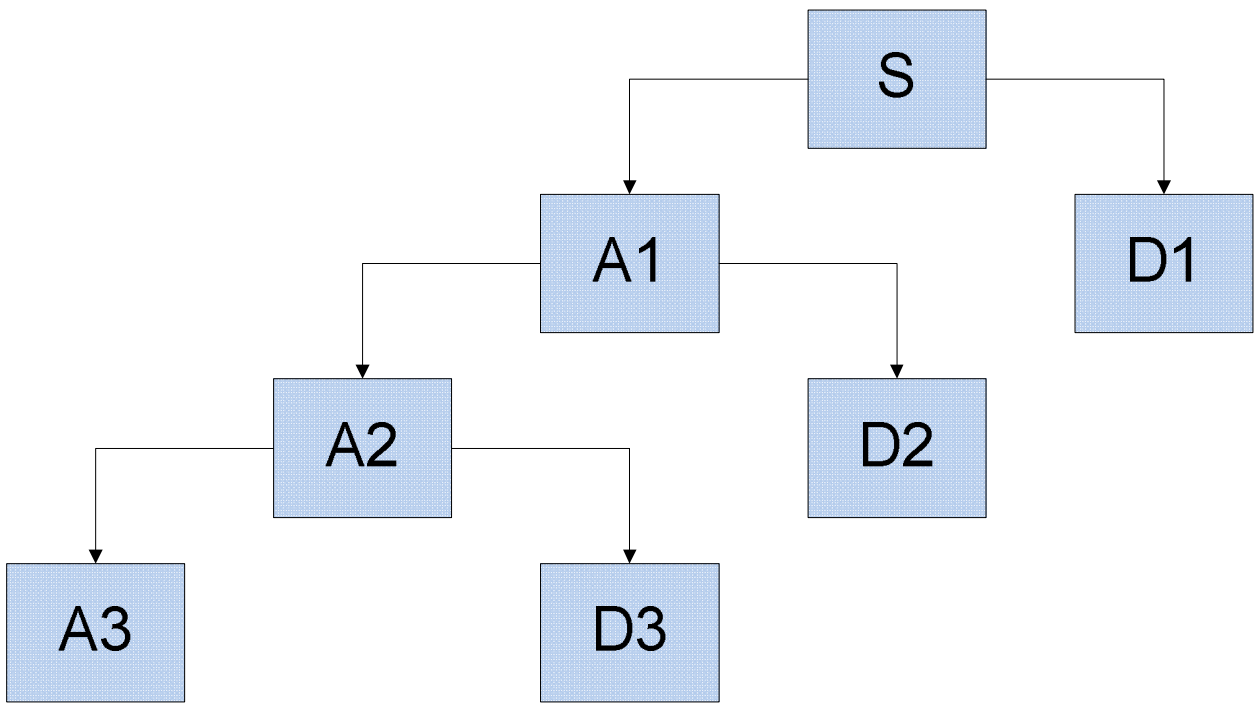

Fig 1 : Wavelet Decomposition Levels

The advantage of wavelet transform is unlike Fourier Transform is not only it gives the information about the frequency spectra of the signal but also provides time localization of it.

## III. Feature Extraction

Feature extraction involves extraction of statistical properties from decomposed signals. Extracted features are used as input for machine learning based classifier.

### A. *Standard Deviation*

It is the amount of deviation from the average value of the signal. Mathematically it can be denoted as

$$\sigma(t_a,t_b) = \left( \int_{t_a}^{t_b} (\mathrm{x(t)} - \bar{x})^2 \, dt \right)^{\frac{1}{2}} \quad \ldots\ldots\ldots\ldots\ldots\ldots(1)$$

where,

$$\bar{x} = \frac{1}{t_a - t_b} \int_{t_a}^{t_b} x(\mathrm{t})\,\mathrm{dt} \quad \ldots\ldots\ldots\ldots\ldots\ldots\ldots\ldots\ldots\ldots\ldots(2)$$

### B. *Mean*

It is the average of the signals that are used.

$$\bar{x} = \frac{1}{t_a - t_b} \int_{t_a}^{t_b} x(\mathrm{t})\,\mathrm{dt} \quad \ldots\ldots\ldots\ldots\ldots\ldots\ldots\ldots(3)$$

### C. *Kurtosis*

It gives the measure of amount of outlier in the signal .

$$K(t_a,t_b) = \int_{t_a}^{t_b} \left( \frac{x(t) - \bar{x}}{\sigma} \right)^4 \quad \ldots\ldots\ldots\ldots\ldots\ldots(4)$$

### D. *Skewness*

It gives the measure of amount of asymmetry in the signal.

$$S(\mathrm{t}_a,\mathrm{t}_b) = \frac{\int_{t_a}^{t_b} (\mathrm{x(t)} - \bar{x})^3}{\sigma^3} \quad \ldots\ldots\ldots\ldots\ldots\ldots(5)$$

### E. *Entropy*

It can be defined as the amount of randomness in the signal that can be measured statistically.

$$ENT = -\sum_{i=1}^{n} P(\mathrm{x}_i)\log_2 P(\mathrm{x}_i) \quad \ldots\ldots\ldots\ldots\ldots(6)$$

## IV. Flowchart of Detection and Classification

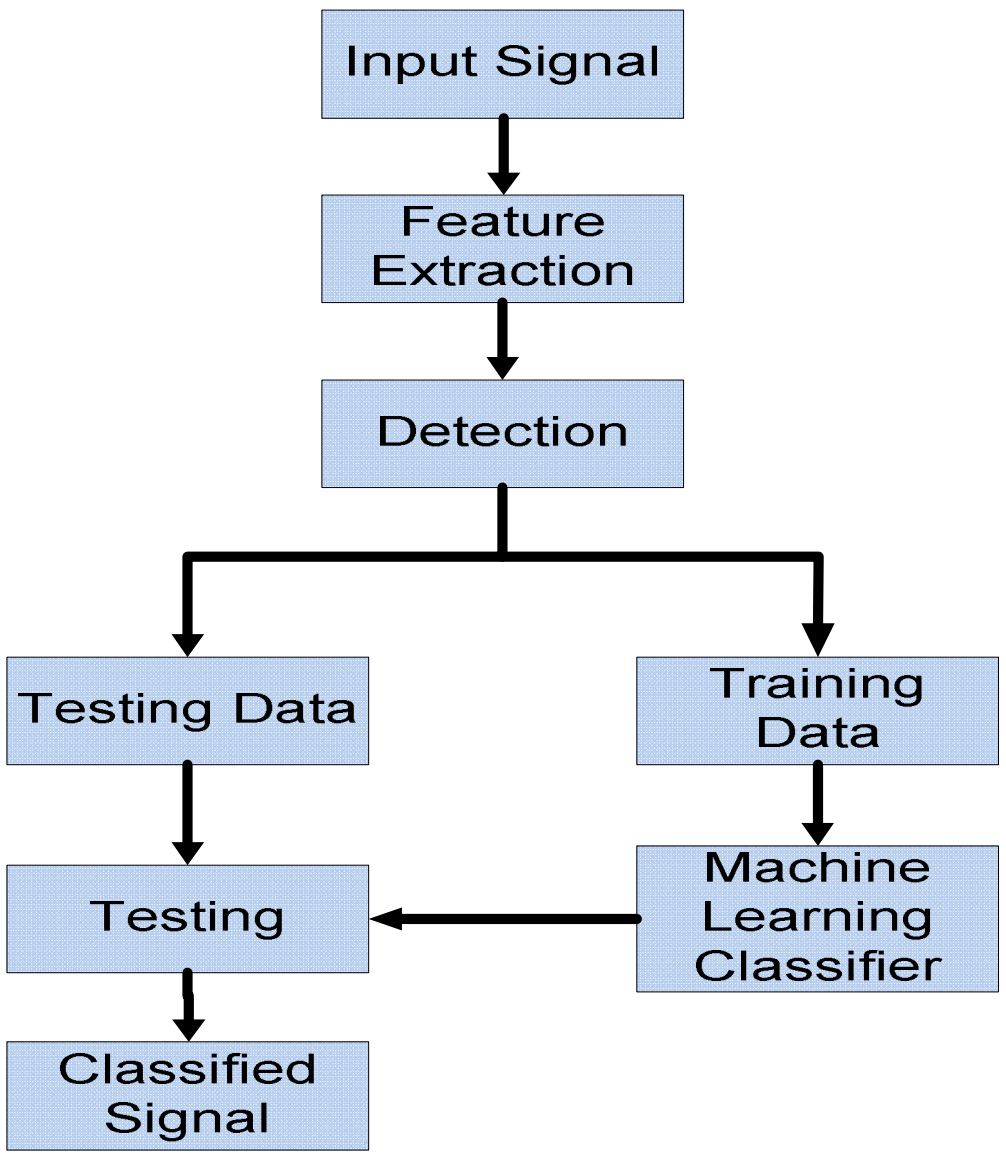

Fig 2 : Flowchart of detection and classification

## V. MACHINE LEARNING BASED CLASSIFIERS

With development in the field of machine learning, multiple powerful classifiers have been developed. In this paper three classifiers are used for classification of power quality events.

### *A. K-Nearest Neighbour (KNN)*

KNN algorithm attempts to find a correlation between the unknown samples and the samples presents in the training set[8]. The numeric features of the training set is denoted by the 'n' dimensions. The algorithm mines for samples that nearby or adjacent to the unknown samples for the given classification process. The Euclidean distance provides information regarding the amount of closeness the samples have and can be mathematically denoted as

$$D(X,Y)=\sqrt{\sum_{i=1}^{n}(x_i-y_i)^2} \quad \text{.......................(7)}$$

where x is the input and y is the target.

The classification of the unknown class is based on the class that occurs the maximum number of times. By setting the value of i=1 the unknown sample is mapped to the nearest neighbor. The uniqueness of this model is it does not create a definite model during the training phase.

### *B. Support Vector Machine (SVM)*

Support vector machine is a structural risk minimization based classifier[9].It is basically a hyperplane dividing different classes of data. The advantage of SVM is its ability to generalize data when transformed to a high dimensional feature space.

The hyperplane is defined as

$$f(\mathrm{s})=\mathrm{w}^T s+b=\sum_{j=1}^{n} w_j s_j+b \quad \text{.............(8)}$$

where w can be defined as the weight vector of n dimension and b is a constrained parameter value.

Magnitude of the hyperplane is defined by w and b values. The optimal hyperplane is a minimization problem denoted as

$$\text{Minimize } \tfrac{1}{2}\|\mathrm{w}\|^{-2}+\mathrm{C}\sum_{\mathrm{i}=1}^{\mathrm{M}}\xi_{\mathrm{i}} \quad \text{............(9)}$$

subject to

$$o_i(\mathrm{w}^T\mathrm{s}+\mathrm{b})\geq 1-\xi_i,\ \text{for i = 1,2,….M} \quad \text{............(10)}$$

The optimal value of b is given as

$$\mathrm{b}^*=-\tfrac{1}{2}\sum_{SVs} o_i\alpha_i^*(\upsilon_1^T\,\mathrm{s}_i+\upsilon_2^T\,\mathrm{s}_i) \quad \text{..............(11)}$$

where $\upsilon_1^T$ and $\upsilon_2^T$ are SVM of class 1 and class 2

The decision set is defined as

$$f(\mathrm{s})=\sum_{SVs}\alpha_i o_i s_i^T s+b^* \quad \text{..............(12)}$$

In this paper for classification, radial basis function is used as kernel.

### *C. Random Forest*

Decision tree is a popular machine learning algorithm useful for classification and regression tasks. It is simple and comparatively easier to use[10]. The issues that plague Decision Tree are bad performance and lack of robustness. In view of such short comings a new technique was developed in which the trees were clustered together with others to form an ensemble followed by a vote for the most popular class labeled forest and the algorithm was named Random Forest. An ensemble method consist of sorting a number of the weak learners and boosting their performance via a voting scheme. The prominent characteristics of Random forest are bootstrap resampling with random feature selection and out of bag error estimation.Suppose there are X inputs where $X=\{x_1,x_2,x_3....x_m\}$ is a m dimensional vector. This input is fed to an ensemble of C trees where trees are denoted by $T_1(X),T_2(X),T_3(X)...T_C(X)$. This results in output Y which is the predicted value generated by an ensemble of trees. After the generation of prediction by each tree the average of the predicted values is taken which is the final predicted value.

## VI. PQ DISTURBANCE GENERATION AND ANALYSIS

For the purpose of detection and classification, a database of different PQ disturbances are generated synthetically using MATLAB. For training of classifier it is advantageous to use synthetic data as real time data collection and usage is extremely tedious and may consist of bad data or data with noise which may distort the classification of accuracy of the classifier. In this paper numerical model of different single and multi stage disturbances are taken from [11]. These mathematical models closely depict the real time PQ events and comply with the standards set by IEEE. While simulating such events a sampling frequency of 3.2 KHz is taken.11 classes of PQ events are taken. The signals are passed through Daubechies4 Wavelet and the entropy, standard deviation, mean, skewness and kurtosis of level 3 detail coefficient are used to train machine learning classifiers. 700 signals of each event at varying amplitude and time durations are generated out of which the classifier is trained using features extracted from 600 signals and tested using features extracted from 100 signals. The inbuilt wavelet transform in MATLAB is used for decomposition of signals.

Different PQ disturbances taken for detection and classification:

| | |
|---|---|
| C1 → Swell with harmonics | C2 → Swell |
| C3 → Spike | C4 → Sag with harmonics |
| C5 → Sag with harmonics | C6 → Oscillatory Transient |
| C7 → Notch | C8 → Interruption with harmonics |
| C9 → Interruption | C10 → Flicker with harmonics |
| C11 → Flicker | |

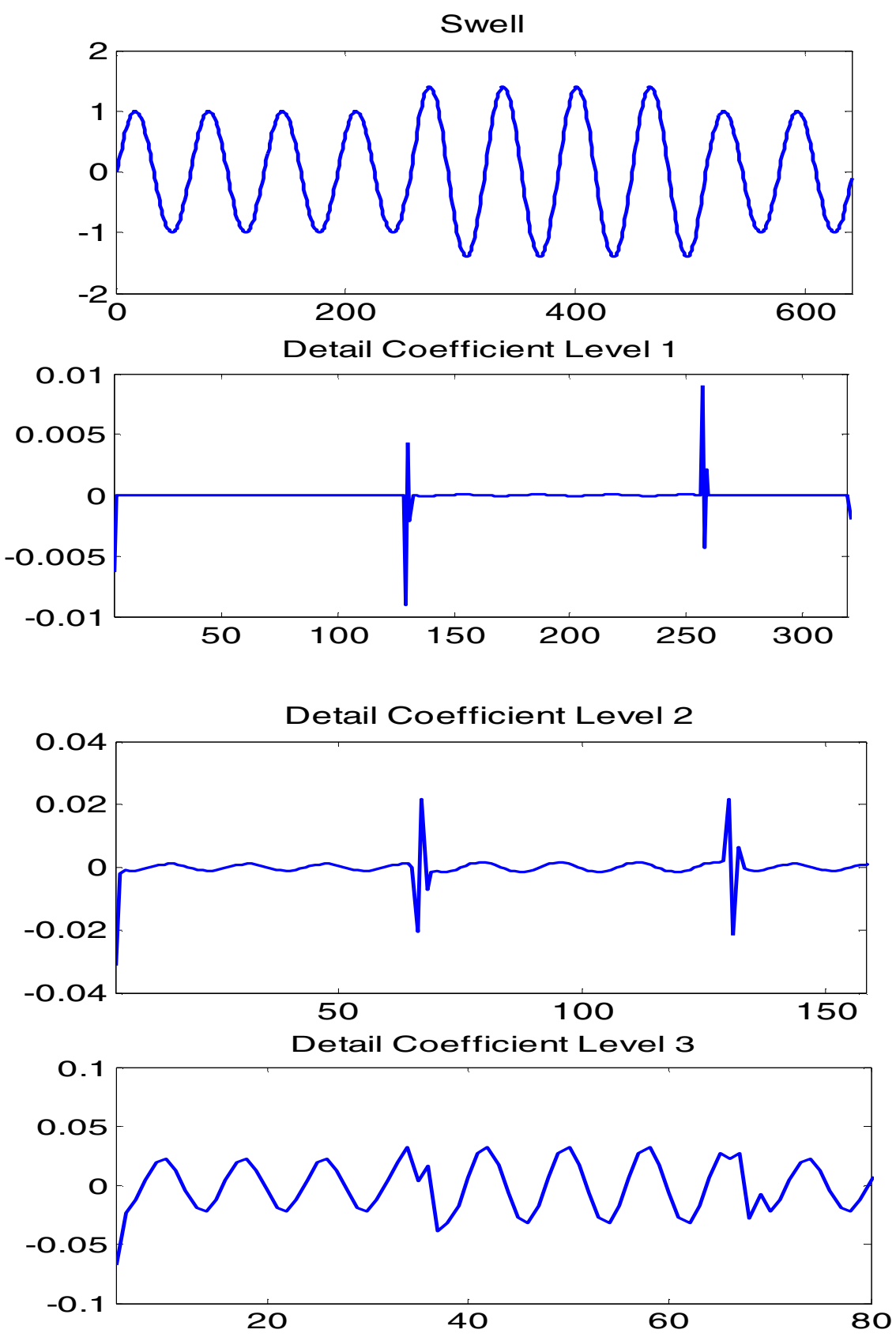

Fig 3: Swell with corresponding detail coefficient level 1,2,3

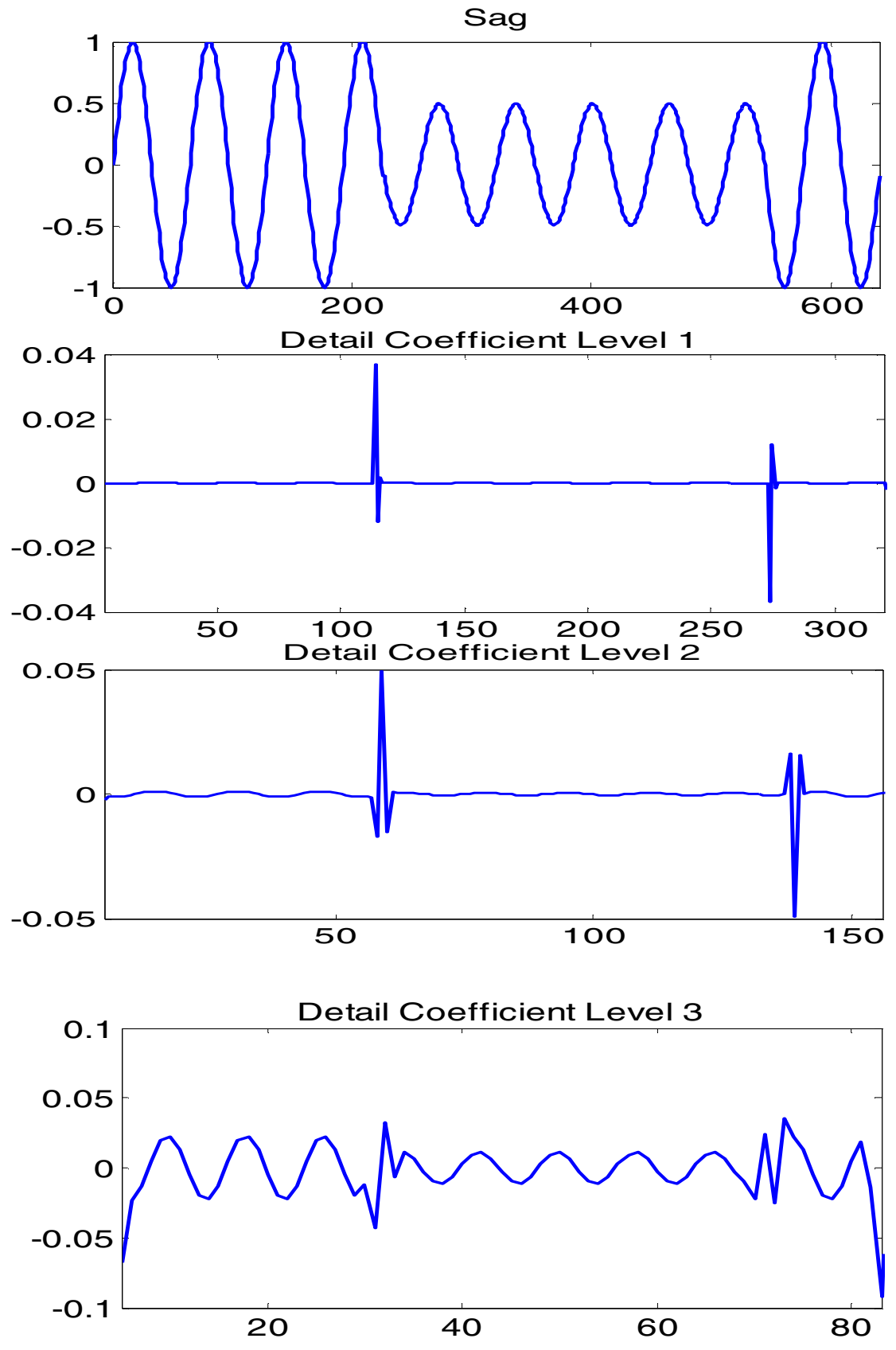

Fig 4: Swell with corresponding detail coefficient level 1,2,3

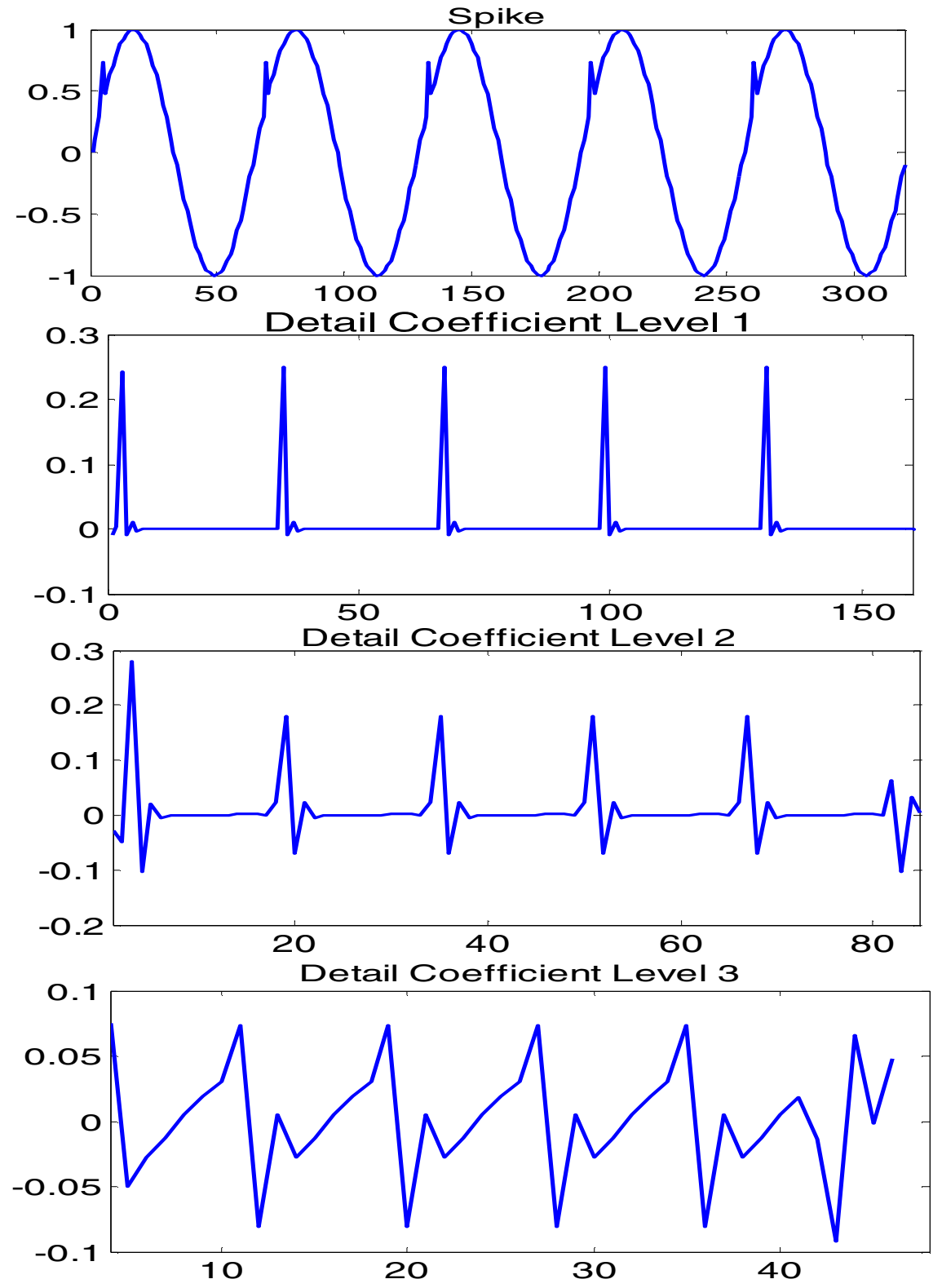

Fig 5: Spike with corresponding detail coefficient level 1,2,3

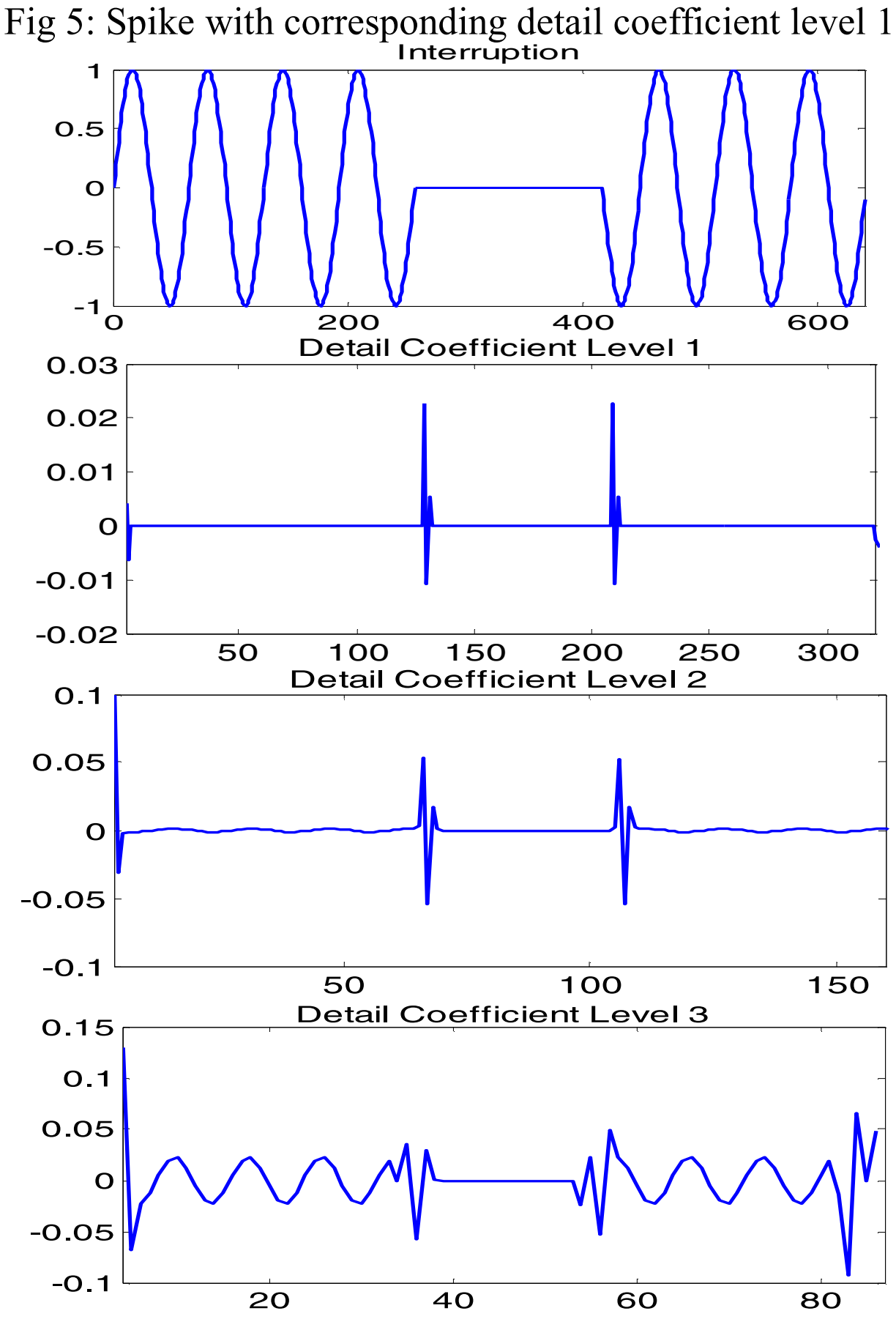

Fig 6: Interruption with corresponding detail coefficient level 1,2,3

## VII. Result Analysis

After feature extraction using wavelet transform, the extracted features are fed to machine learning classifiers. After training the classifier is tested on unknown data. The confusion matrix of machine learning classifiers along with their testing accuracy is illustrated below:

Table 1:Confusion Matrix of KNN

| PQD | C1 | C2 | C3 | C4 | C5 | C6 | C7 | C8 | C9 | C10 | C11 |
|---|---|---|---|---|---|---|---|---|---|---|---|
| 1 | **93** | 0 | 0 | 0 | 0 | 0 | 0 | 0 | 0 | 7 | 0 |
| 2 | 0 | **87** | 0 | 0 | 7 | 0 | 0 | 0 | 0 | 0 | 6 |
| 3 | 0 | 0 | **100** | 0 | 0 | 0 | 0 | 0 | 0 | 0 | 0 |
| 4 | 0 | 0 | 0 | **100** | 0 | 0 | 0 | 0 | 0 | 0 | 0 |
| 5 | 0 | 40 | 0 | 0 | **60** | 0 | 0 | 0 | 0 | 0 | 0 |
| 6 | 0 | 0 | 0 | 0 | 0 | **100** | 0 | 0 | 0 | 0 | 0 |
| 7 | 0 | 0 | 0 | 0 | 0 | 0 | **100** | 0 | 0 | 0 | 0 |
| 8 | 0 | 0 | 0 | 0 | 0 | 0 | 0 | **100** | 0 | 0 | 0 |
| 9 | 0 | 0 | 0 | 0 | 0 | 0 | 0 | 0 | **100** | 0 | 0 |
| 10 | 2 | 0 | 0 | 0 | 0 | 0 | 0 | 0 | 0 | **98** | 0 |
| 11 | 0 | 8 | 0 | 0 | 36 | 0 | 0 | 0 | 0 | 0 | **66** |
| Overall Accuracy = 90.36% | | | | | | | | | | | |

Table 2:Confusion Matrix of SVM

| PQD | C1 | C 2 | C3 | C4 | C5 | C6 | C7 | C8 | C9 | C10 | C11 |
|---|---|---|---|---|---|---|---|---|---|---|---|
| 1 | **100** | 0 | 0 | 0 | 0 | 0 | 0 | 0 | 0 | 0 | 0 |
| 2 | 0 | **98** | 0 | 0 | 2 | 0 | 0 | 0 | 0 | 0 | 0 |
| 3 | 0 | 0 | **100** | 0 | 0 | 0 | 0 | 0 | 0 | 0 | 0 |
| 4 | 0 | 0 | 0 | **100** | 0 | 0 | 0 | 0 | 0 | 0 | 0 |
| 5 | 0 | 0 | 0 | 0 | **100** | 0 | 0 | 0 | 0 | 0 | 0 |
| 6 | 0 | 0 | 0 | 0 | 0 | **100** | 0 | 0 | 0 | 0 | 0 |
| 7 | 0 | 0 | 0 | 0 | 0 | 0 | **100** | 0 | 0 | 0 | 0 |
| 8 | 0 | 0 | 0 | 0 | 0 | 0 | 0 | **100** | 0 | 0 | 0 |
| 9 | 0 | 0 | 0 | 0 | 16 | 0 | 0 | 0 | **86** | 0 | 0 |
| 10 | 7 | 0 | 0 | 0 | 0 | 0 | 0 | 0 | 0 | **93** | 0 |
| 11 | 0 | 0 | 0 | 0 | 0 | 0 | 0 | 0 | 0 | 0 | **100** |
| Overall Accuracy = 97.72% | | | | | | | | | | | |

Table 3:Confusion Matrix of Random Forest

| PQD | C1 | C2 | C3 | C4 | C5 | C6 | C7 | C8 | C9 | C10 | C11 |
|---|---|---|---|---|---|---|---|---|---|---|---|
| 1 | **100** | 0 | 0 | 0 | 0 | 0 | 0 | 0 | 0 | 0 | 0 |
| 2 | 0 | **100** | 0 | 0 | 0 | 0 | 0 | 0 | 0 | 0 | 0 |
| 3 | 0 | 0 | **100** | 0 | 0 | 0 | 0 | 0 | 0 | 0 | 0 |
| 4 | 0 | 0 | 0 | **100** | 0 | 0 | 0 | 0 | 0 | 0 | 0 |
| 5 | 0 | 3 | 0 | 0 | **97** | 0 | 0 | 0 | 0 | 0 | 0 |
| 6 | 0 | 0 | 0 | 0 | 0 | **100** | 0 | 0 | 0 | 0 | 0 |
| 7 | 0 | 0 | 0 | 0 | 0 | 0 | **100** | 0 | 0 | 0 | 0 |
| 8 | 0 | 0 | 0 | 0 | 0 | 0 | 0 | **100** | 0 | 0 | 0 |
| 9 | 0 | 0 | 0 | 0 | 0 | 0 | 0 | 0 | **100** | 0 | 0 |
| 10 | 2 | 0 | 0 | 0 | 0 | 0 | 0 | 0 | 0 | **98** | 0 |
| 11 | 0 | 0 | 0 | 0 | 0 | 0 | 0 | 0 | 0 | 0 | **100** |
| Overall Accuracy = **99.54%** | | | | | | | | | | | |

From the confusion matrix (Table 1,2,3) of KNN, SVM and Random Forest it is seen that the overall classification efficiency of KNN is 90.36%, SVM is 97.72% and Random Forest is 99.54%. This clearly shows the superior accuracy of Random Forest in classification of PQ events.

## VIII. Conclusion

In this paper wavelet transform is used to extract features from power quality disturbance. KNN, SVM and Random Forest algorithm are used for automatic disturbance classification of power quality events using feature set obtained from wavelet transform. The classification results of the confusion matrix clearly illustrate the superior accuracy of Random Forest based classifier over SVM and KNN.